\documentclass[twocolumn,prb]{revtex4}
\usepackage{amsfonts}
\usepackage[T1]{fontenc}
\usepackage{amsmath,amsbsy,amssymb,graphicx}
\usepackage{times}

\let\mathbf=\boldsymbol

\begin{document}

\title{Electric circuit simulations of $n$th-Chern insulators in $2n$-dimensional space \\
and their non-Hermitian generalizations for arbitral $n$}
\author{Motohiko Ezawa}
\affiliation{Department of Applied Physics, University of Tokyo, Hongo 7-3-1, 113-8656,
Japan}

\begin{abstract}
We show that topological phases of the Dirac system in arbitral even
dimensional space are simulated by $LC$ electric circuits with operational
amplifiers. The lattice Hamiltonian for the hypercubic lattice in $2n$
dimensional space is characterized by the $n$-th Chern number. The boundary
state is described by the Weyl theory in $2n-1$ dimensional space. They are
well observed by measuring the admittance spectrum. They are different from
the disentangled $n$-th Chern insulators previously reported, where the $n$-th Chern 
number is a product of the first Chern numbers. The results are
extended to non-Hermitian systems with complex Dirac masses. The
non-Hermitian $n$-th Chern number remains to be quantized for the complex
Dirac mass.
\end{abstract}

\maketitle

\section{Introduction}

Topological insulators were originally discovered in condensed matter
physics. Now the notion has been extended to cold atoms\cite{Gold}, photonics%
\cite{TopoPhoto2014,TopoPhoto} and acoustics\cite{TopoAco,Xue,Alex,WZhu}. A
recent finding is that it is also applicable to electric circuits\cite%
{ComPhys,TECNature}. Various topological phases have been simulated by
electric circuits\cite%
{ComPhys,TECNature,Garcia,Hel,Lu,EzawaTEC,Hofmann,EzawaLCR,EzawaChern,EzawaMajo}. 
In this context, non-Hermitian topological systems are among hottest
topics of artificial topological systems\cite%
{Zeu,Konotop,Mark,Scho,Pan,Hoda,NoriSOTI,EzawaLCR}. Nevertheless, all these
phases are also realizable in condensed matter in principle. It is an
interesting problem to explore exotic topological phases which would never
exist in condensed matter.

An obvious example is the topological phase in the spatial $d$ dimensions 
($d $Ds) with $d\geq 4$. The four-dimensional (4D) quantum Hall effect is one
of them\cite{4D}. It is simulated via 2D quasi-crystals\cite{Kraus} or
topological pumpings in photonic systems\cite{Zil,Ozawa} and optical lattices%
\cite{Price,Lohse}. They are characterized by the second-Chern number. In
the same way, the 6D quantum Hall effect is realized in 3D topological
pumpings\cite{6D}. However, these phases are disentangled, where they are
decomposed into two or three independent copies of 2D quantum Hall
insulators. Accordingly, the second- or third-Chern numbers become products
of the first-Chern numbers. We seek genuine higher-dimensional topological
phases, which cannot be decomposed into lower-dimensional topological phases.

In this paper, we show that topological phases of the Dirac system in any
even dimensional space are simulated by $LC$ electric circuits with
operational amplifiers. Especially, we construct topological phases
characterized by the $n$-th Chern number in $2n$Ds for arbitral $n$. We
start with the Dirac Hamiltonian in higher dimensions. Then, we analyze
lattice models defined on the hypercubic lattice. The $n$-th Chern number is
analytically calculated since it is determined by the Dirac theory at the 
$2^{2n}$ high-symmetry points. Then we show that the boundary states are
described by the Weyl theory in (2n-1)D. They are manifested by calculating
the density of states (DOS), which is proportional to $|E|^{2n-2}$. Finally,
we point out that they are well signaled by admittance spectrum.

\section{Model Hamiltonian}

The Dirac Hamiltonian in $2n$D space is given by

\begin{equation}
H_{2n}=\int d^{2n}k\left[ \psi ^{\dagger }\left( x\right) \Gamma _{j}\left(
-i\partial _{j}\right) \psi \left( x\right) +M\psi ^{\dagger }\Gamma
_{0}\psi \right] ,  \label{Dirac}
\end{equation}%
where $\Gamma _{\mu }$ with $j=1,2,\cdots ,2n$ and $\mu =0,1,2,\cdots ,2n$
are the gamma matrices satisfying the Clifford algebra $\left\{ \Gamma _{\mu
},\Gamma _{\nu }\right\} =2\delta _{\mu \nu }I$, while $M$ is the Dirac
mass. The system becomes non-Hermitian when the mass $M$\ is taken
to be complex. All the analysis is valid both for Hermitian and
non-Hermitian systems.

The gamma matrix $\Gamma _{\mu }^{2n}$ in $2n$Ds has a $2^{n}$ dimensional
representation, which is recursively defined by%
\begin{eqnarray}
\Gamma _{0}^{2n} &=&\sigma _{x}\otimes \mathbb{I}^{2n-2},\quad \Gamma
_{2n}^{2n}=\sigma _{z}\otimes \mathbb{I}^{2n-2},  \notag \\
\Gamma _{j}^{2n} &=&\sigma _{y}\otimes \Gamma _{j}^{2n-2},\text{ }
\end{eqnarray}%
for $1\leq j\leq 2n-1$, and $\Gamma _{0}^{2}=\sigma _{z}$, $\Gamma
_{1}^{2}=\sigma _{x}$, $\Gamma _{2}^{2}=\sigma _{y}$.

The corresponding lattice model is\cite{Golt,TQFT},%
\begin{equation}
H_{2n}\left( k\right) =\sum_{\mu =0}^{2n}\psi ^{\dagger }\left( k\right)
d^{\mu }\Gamma _{\mu }\psi \left( k\right) ,  \label{Wilson}
\end{equation}%
where%
\begin{equation}
d^{0}=m+t\sum_{j=1}^{2n}\cos k_{j},\qquad d^{j}=\lambda \sin k_{j},
\label{ParamD}
\end{equation}%
with the on-site potential $m$, the hopping amplitude $t$ and the
spin-orbital interaction $\lambda $. The energy spectrum is given by $E=\pm 
\sqrt{\sum_{\mu =0}^{2n}\left( d^{\mu }\right) ^{2}}$.

\begin{figure}[t]
\centerline{\includegraphics[width=0.48\textwidth]{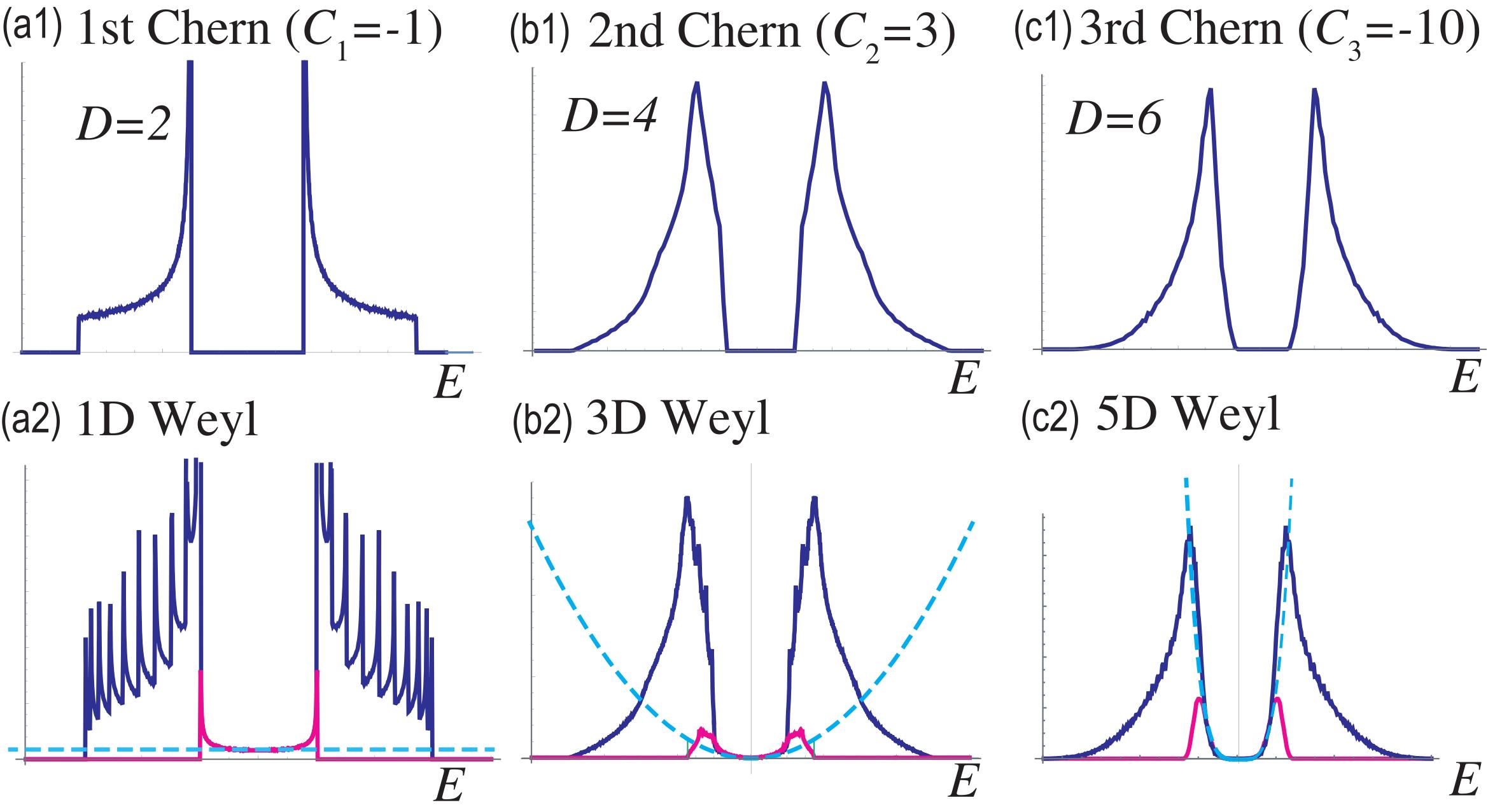}}
\caption{(a1)--(c1) DOS of the bulk in $n$-th Chern insulators for the
Hermitian system as a function of the energy $E$ for (a) $D=2$ with 
$C_{1}=-1 $, (b) $D=4$ with $C_{2}=3$, and (c) $D=6$ with $C_{3}=-10$. There
are finite gaps, dictating that the systems are insulators. (a2)--(c2) DOS
of the boundary states as a function of the energy $E$. The dark blue curves
represent the DOS for all bands numerically obtained, the magenta curves
represent the DOS for the energy (\protect\ref{ESinK}) analytically
obtained, and the dotted cyan curves represent the DOS for the energy 
(\protect\ref{Ek}) analytically obtained. These numerical results are well
fitted by analytical results of the Weyl theory in the vicinity of the Fermi
energy. We have set $t=\protect\lambda =m=1$. }
\label{FigDOS}
\end{figure}

First, we show the bulk DOS of the Hermitian system in Fig.\ref{FigDOS}(a1)-(c1). 
The system is an insulator due to a finite gap, which indicates
that the system is an insulator.

\begin{figure}[t]
\centerline{\includegraphics[width=0.48\textwidth]{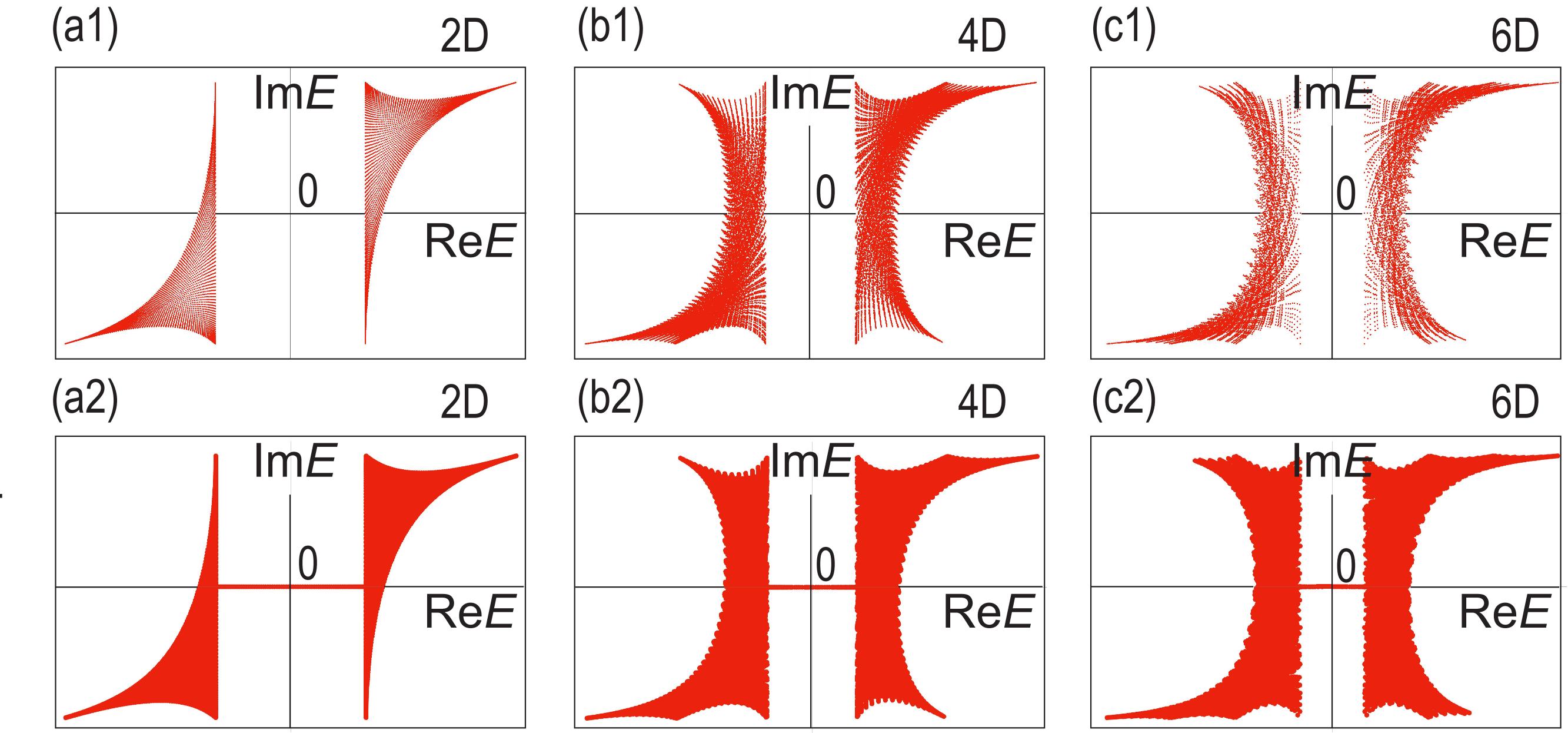}}
\caption{ Energy spectrum in the Re$E$-Im$E$ plane for (a) $D=2$ ($L=100$),
(b) $D=4$ ($L=25$) and (c) $D=6$ ($L=25$), where $L$ is the length of the hypercube. (a1)--(c1) There are gaps in the
bulk spectrum, showing that the system is an insulator. (a2)--(2) There are
boundary states connecting two bulk spectra, which is on the Im$E=0$ line.
They are Weyl boundary states. We have set $t=\protect\lambda =1$ and $m=1+0.1i$.}
\label{FigLineGap}
\end{figure}

Next, we show the energy spectrum in the Re$E$-Im$E$ plane\cite{Fu} for the
non-Hermitian system in Fig.\ref{FigLineGap}(a1)-(c1). The system remains to
be an insulator. The two bulk spectra are separated by a line given by Re$E=0 $. 
This structure is called a line gap\cite{KawabataST}, which is a
generalization of an insulator to the non-Hermitian system.

The gap closes at the $2^{2n}$ high-symmetry points $\mathbf{K}=\left(K_{1},K_{2},\cdots ,K_{2n}\right) $, where $K_{j}=0,\pi $. 
By evaluating (\ref{ParamD}) at these points, we obtain a set of $2^{2n}$ Dirac
Hamiltonians (\ref{Wilson}) together with%
\begin{eqnarray}
d^{0}\left( \mathbf{K}\right) &=&m+t\sum_{j=1}^{2n}\left( -1\right)
^{K_{j}/\pi }\equiv \bar{m},  \label{d0} \\
d^{j}\left( \mathbf{K}\right) &=&\left( -1\right) ^{K_{j}/\pi }\lambda k_{j},
\label{dj}
\end{eqnarray}%
where $d^{0}\left( \mathbf{K}\right) $ is the Dirac mass at the point 
$\mathbf{K}$. We define the sign of the Dirac Hamiltonian by $\eta \left( 
\mathbf{K}\right) \equiv \left( -1\right) ^{\sum K_{j}/\pi }=\pm 1$.

\section{Non-Hermitian $n$-th Chern number}

In the classification table, the topological insulator without any symmetry
is classified by the $\mathbb{Z}$ index for both of the Hermitian\cite%
{Schny,Ryu} and non-Hermitian systems\cite{KawabataST}. In the case of the
Hermitian system, the topological number is the $n$-th Chern number. It is
also applicable to the non-Hermitian system by using the following
definition.

As a generalization of the non-Hermitian "first" Chern number\cite%
{Kohmoto,Fu,Yao2,Kunst,KawabataChern,Philip,Chen,EzawaChern}, we define the
non-Hermitian $n$-th Chern number as%
\begin{equation}
C_{n}=\frac{1}{\left( 2\pi \right) ^{n}}\int \frac{1}{n!}\bigwedge%
\limits^{n}Fd^{2n}k,  \label{Chern}
\end{equation}%
where $F$ is the non-Hermitian non-Abelian Berry curvature,%
\begin{equation}
F_{ij}^{\alpha \beta }=\partial _{i}A_{j}^{\alpha \beta }-\partial
_{j}A_{i}^{\alpha \beta }+i\left[ A_{i},A_{j}\right] ^{\alpha \beta },
\end{equation}%
with the non-Hermitian Berry connection\cite{Kohmoto,Zhu,Yin,Lieu} 
\begin{equation}
A_{i}^{\alpha \beta }\left( k\right) =-i\left\langle \psi _{\alpha }^{\text{L}}\left( k\right) 
\right\vert \partial _{k_{i}}\left\vert \psi _{\beta }^{\text{R}}\left( k\right) \right\rangle .
\end{equation}%
Here, $\left\vert \psi _{\beta }^{\text{R}}\left( k\right) \right\rangle $
is the right-eigen states $H\left\vert \psi _{\beta }^{\text{R}}\left(
k\right) \right\rangle =E\left\vert \psi _{\beta }^{\text{R}}\left( k\right)
\right\rangle $, and $\left\vert \psi _{\beta }^{\text{L}}\left( k\right)
\right\rangle $ is the right-eigen states $H^{\dagger }\left\vert \psi
_{\beta }^{\text{L}}\left( k\right) \right\rangle =E^{\ast }\left\vert \psi
_{\beta }^{\text{L}}\left( k\right) \right\rangle $.

Substituting (\ref{d0}) and (\ref{dj}) into (\ref{Chern}), the $n$-th Chern
number is given by 
\begin{equation}
C_{n}\left( K_{\mu }\right) =\eta \frac{\left( 4n-1\right) !!}{2^{n-1}\left(
2n-1\right) !}\int d^{2n}k\frac{\bar{m}}{\left( k^{2}+\bar{m}^{2}\right)
^{\left( 2n+1\right) /2}},
\end{equation}%
which is explicitly calculated as%
\begin{equation}
C_{n}\left( K_{\mu }\right) =\eta \frac{\bar{m}}{2\sqrt{\bar{m}^{2}}}.
\label{ChernK}
\end{equation}%
It reads $C_{n}\left( K_{\mu }\right) =\eta /2$ for Re$\left[ \bar{m}\right]
>0$ and $C_{n}\left( K_{\mu }\right) =-\eta /2$ for Re$\left[ \bar{m}\right]
<0$. It is quantized even for a complex mass.

The topological phase boundaries are determined by the massless condition 
$\bar{m}\left( K_{\mu }\right) =0$, which is independent of the value of the
spin-orbital interaction $\lambda $.

\begin{figure}[t]
\centerline{\includegraphics[width=0.48\textwidth]{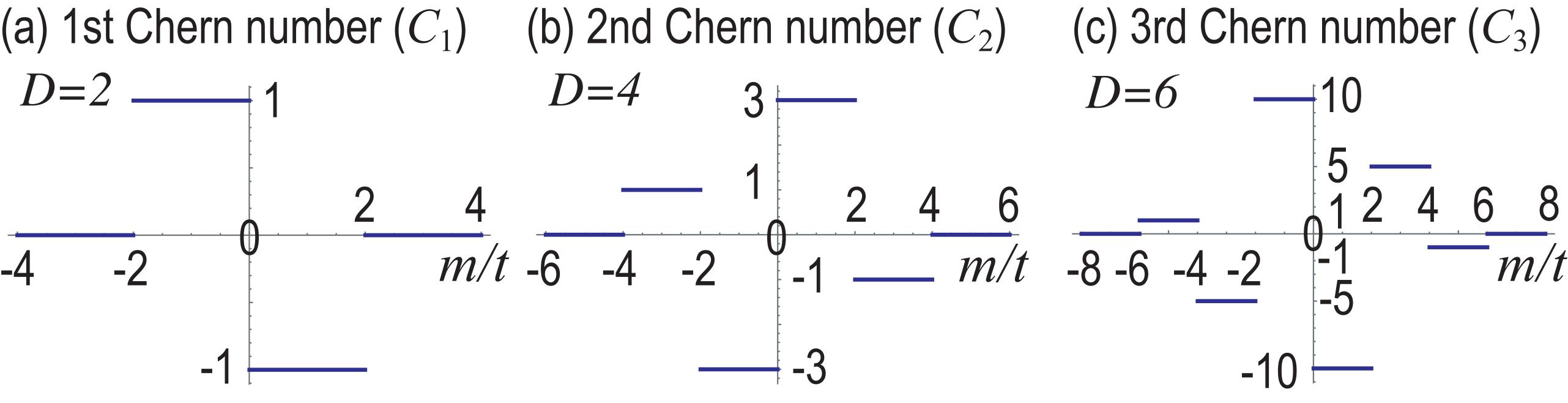}}
\caption{ $n$-th Chern number as a function of $m$ for (a) $D=2$, (b) $D=4$
and (c) $D=6$. We have set $\protect\lambda =1$.}
\label{FigChernM}
\end{figure}

Then the total $n$-th Chern number for the lattice Hamiltonian is 
\begin{equation}
C_{n}=\sum_{K_{j}}C_{n}\left( K_{j}\right) ,
\end{equation}%
where the summation is taken over all the highest symmetry points. We show
the $n$-th Chern number as a function of $m/t$ in Fig.\ref{FigChernM}. It is
always quantized and jumps when the mass becomes zero.

\section{Weyl Boundary states}

The boundary states emerge in a topological phase of an insulator. Let us
review the case of 2Ds, where we analyze a nanoribbon with a finite width
along the $x_{2}$ axis. First of all, the dynamical degree of freedom with
respect to $x_{2}$ is frozen by the Jackiw-Rebbi solution of the Dirac
equation\cite{JR}, where the mass has a spatial dependence. On the other
hand, the lattice structure along the $x_{1}$ axis allows us to introduce
the crystal momentum $k_{1}$. Consequently, the dynamical degree of freedom
is carried solely by the momentum $k_{1}$ along the edge.

Similarly in 2$n$Ds, we analyze a nanostructure with a finite thickness
along the $x_{2n}$ axis, while the other directions are periodic. We
numerically calculate the energy spectrum of the non-Hermitian system in the
Re$E$-Im$E$ plane as in Fig.\ref{FigLineGap}(a2)--(c2). There are lines
connecting two separated bulk spectra along the Im$E=0$ line in topological
phases. They are Weyl boundary states, which remains to be real even for the
non-Hermitian system. It results in the emergence of a zero-energy solution,
which is the Jackiw-Rebbi solution as we now demonstrate.

The Dirac equation describing the boundary states is given by 
\begin{align}
(\bar{m}\left( x_{2n}\right) \sigma _{x}\otimes \mathbb{I}^{2n-2}& -i\lambda
\partial _{x_{2n}}\sigma _{y}\otimes \mathbb{I}^{2n-2} & &  \notag \\
& +\sum_{j=1}^{2n-1}k_{j}\sigma _{z}\otimes \Gamma _{j-1})\psi = & & 0,
\end{align}%
where $\bar{m}\left( x_{2n}\right) $ is a spatially dependent mass, 
\begin{equation}
\bar{m}\left( x_{2n}\right) =m_{0}\tanh \frac{x_{2n}}{\xi },
\label{DiracMass}
\end{equation}%
with $\xi $ the penetration depth. We seek a zero-energy solution by solving%
\begin{equation}
\left( 
\begin{array}{cc}
0 & \bar{m}\left( x_{2n}\right) -\lambda \partial _{x_{2n}} \\ 
\bar{m}\left( x_{2n}\right) +\lambda \partial _{x_{2n}} & 0%
\end{array}%
\right) \left( 
\begin{array}{c}
\psi _{A} \\ 
\psi _{B}%
\end{array}%
\right) =0.
\end{equation}%
It is exactly given by the Jackiw-Rebbi solution,%
\begin{eqnarray}
\psi _{A} &\propto &\exp \left[ -\frac{1}{\lambda }\int \bar{m}\left(
x_{2n}\right) dx_{2n}\right] , \\
\psi _{B} &\propto &\exp \left[ \frac{1}{\lambda }\int \bar{m}\left(
x_{2n}\right) dx_{2n}\right] .
\end{eqnarray}%
The zero-energy solutions exist when the integrands converge. We have shown
that the Jackiw-Rebbi solution follows even in the non-Hermitian system.

\begin{figure}[t]
\centerline{\includegraphics[width=0.28\textwidth]{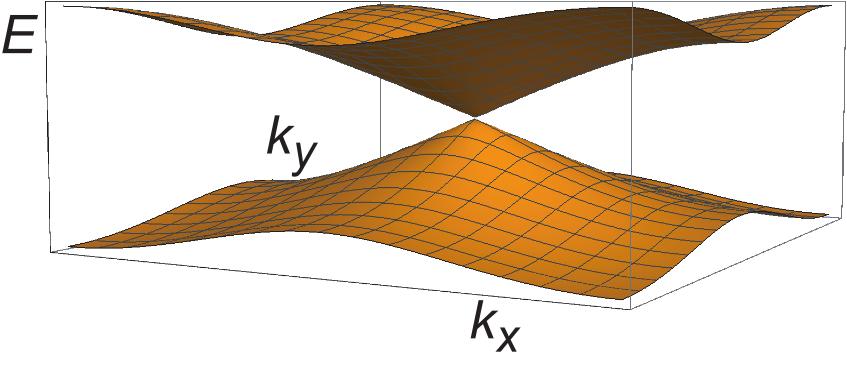}}
\caption{Cone structure of the Weyl boundary state. The vertical axis is $E$, 
while the horizontal axes are $k_{x}$ and $k_{y}$. We have set the other
momenta are zero, $k_{j}=0$ for $3\leq j\leq 2n-1$. It does not depend on the
dimension. The parameters are the same as Fig.\protect\ref{FigLineGap}.}
\label{FigWeyl}
\end{figure}

Since the dynamical degree of freedom with respect to $x_{2n}$ is frozen,
the dimensional reduction occurs from $2n$Ds to ($2n-1$)Ds, where the
dynamical degrees of freedom are crystal momenta $k_{j}$ with $1\leq j\leq
2n-1$. The zero-energy solution is localized at the boundary, where the mass
becomes zero. Hence, the boundary Hamiltonian is described by the Weyl
Hamiltonian in ($2n-1$)Ds,%
\begin{equation}
H_{2n-1}\left( k\right) =\sum_{j=1}^{2n-1}\psi ^{\dagger }\left( k\right)
d^{j}\Gamma _{j}\psi \left( k\right) ,
\end{equation}%
whose energy spectrum is given by%
\begin{equation}
E\left( k\right) =\pm \left\vert \lambda \right\vert \sqrt{\sum_{j=1}^{2n-1}\sin ^{2}k_{j}}.  \label{ESinK}
\end{equation}%
In the vicinity of the Fermi level, the energy dispersion is linear, 
\begin{equation}
E\left( k\right) =\pm \left\vert \lambda k\right\vert .  \label{Ek}
\end{equation}%
We show the boundary states numerically obtained as a function of $k_{x}$
and $k_{y}$ for the $n$-th Chern insulator based on the tight-binding
Hamiltonian (\ref{Wilson}) in Fig.\ref{FigWeyl}, where we have set $k_{j}=0$
for $3\leq j\leq 2n-1$. It is well fitted by (\ref{ESinK}).

The DOS is defined by 
\begin{equation}
\rho \left( E\right) =\int \delta \left( E-\lambda \sqrt{\sum_{j=1}^{2n-1}\sin ^{2}k_{j}}\right) d^{2n-1}k,
\end{equation}%
which is proportional to 
\begin{equation}
\rho \left( E\right) \propto \lambda \left\vert E\right\vert ^{2n-2}
\label{RhoE}
\end{equation}%
in the vicinity of the Fermi level.

We numerically calculate the DOS of the thin film in the case of the
Hermitian system in Fig.\ref{FigDOS}(a2)-(c2). In the vicinity of the Fermi
level, the DOS of the total band is well fitted by the DOS determined by the
boundary lattice Hamiltonian (\ref{ESinK}) and the DOS (\ref{RhoE}) of the
Weyl Hamiltonian. The DOS of the Hamiltonian is observable by measuring
admittance, as we shall see later. Thus, it is an experimental signature to
detect higher-dimensional Weyl boundary states. 

\section{Disentangled $n$-th Chern insulators}

We comment on a trivial construction of the $n$-th Chern insulators, which
are mainly studied in previous literature\cite{Kraus,Zil,Ozawa,Price,Lohse}.
We construct a Hamiltonian by a direct product of the first-Chern insulator
in 2Ds as%
\begin{equation}
H_{2n}\left( k\right) =\bigotimes\limits^{n}H_{2}\left( k\right) .
\end{equation}%
Then the $n$-th Chern number is decomposed into%
\begin{equation}
C_{n}=\left( \frac{1}{2\pi }\int d^{2}kF\right) ^{n}=\prod\limits^{n}C_{1}.
\end{equation}%
It is also an $n$-th Chern insulator. However, it is disentangled and a
trivial example since it is essentially a first-Chern insulator.

\begin{figure}[t]
\centerline{\includegraphics[width=0.48\textwidth]{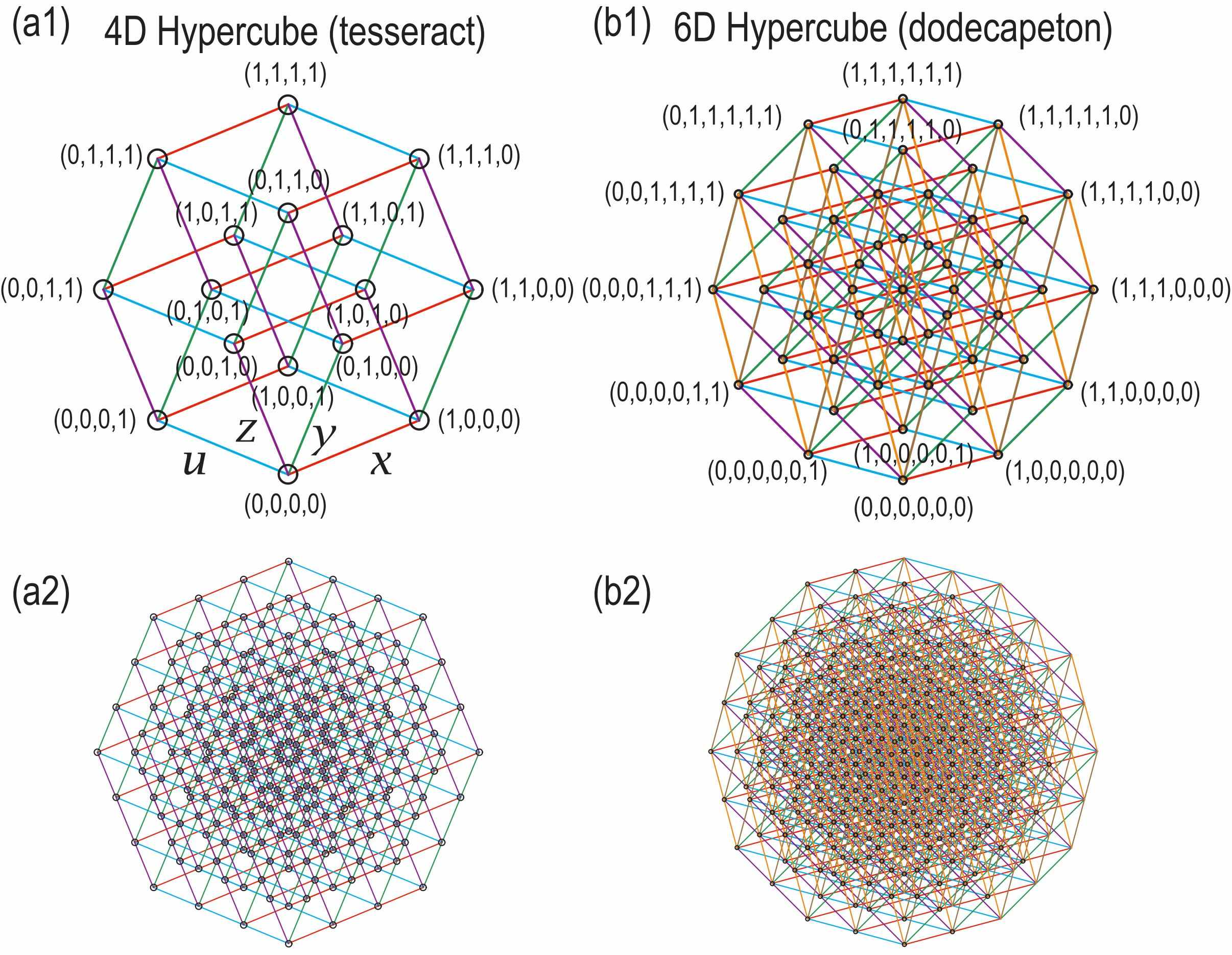}}
\caption{Projection of the hypercube onto the 2D plane. (a) 4D hypercube and
(b) 6D hypercube. Circles represent the unit cell shown in Fig.\protect\ref{FigCircuit}, 
and colored links represent the conection between two unit
cells as in Fig.\protect\ref{FigCircuit}. (a1) Electric circuit for the 4D
hypercube with $L=1$. (a2) The one for the 4D hypercube with $L=3$. (b1) The
one for the 6D hypercube with $L=1$. (b1) The one for the 6D hypercube with $L=2$. 
The coordinates are shown in (a1) and (b1).}
\label{FigProject}
\end{figure}

\section{Electric circuit realization}

We next explain how to simulate the above model in higher dimensions by
electric circuits. First, we note that a hypercubic lattice in any dimensions
can be projected to a 2D plane as illustrated in Fig.\ref{FigProject}. The
lattice points of a hypercube are projected to different positions in the 2D
plane. Although the links cross each other, they can be avoided by using a
bridge structure of wiring. We consider a hypercubic lattice structure in $2n $Ds 
whose unit cell contains $2^{2n}$ sites, which is the dimension of
the $\Gamma $ matrix.

Let us explain an instance of the 4D lattice, i.e., the case of $n=2$. A
single cube ($L=1$) contains $16$ sites as in Fig.\ref{FigProject}(a1). A
pair of two sites are connected by a colored link, which is parallel to one
of the four axes, i.e., the $x$-axis (red), the $y$-axis (green), 
the $z$-axis (purple) and the $u$-axis (cyan). A lattice with size $L=3$ is
illustrated in Fig.\ref{FigProject}(a2), which is obtained by dividing an
edge of a single cube into $L=3$ pieces.

\begin{figure}[t]
\centerline{\includegraphics[width=0.48\textwidth]{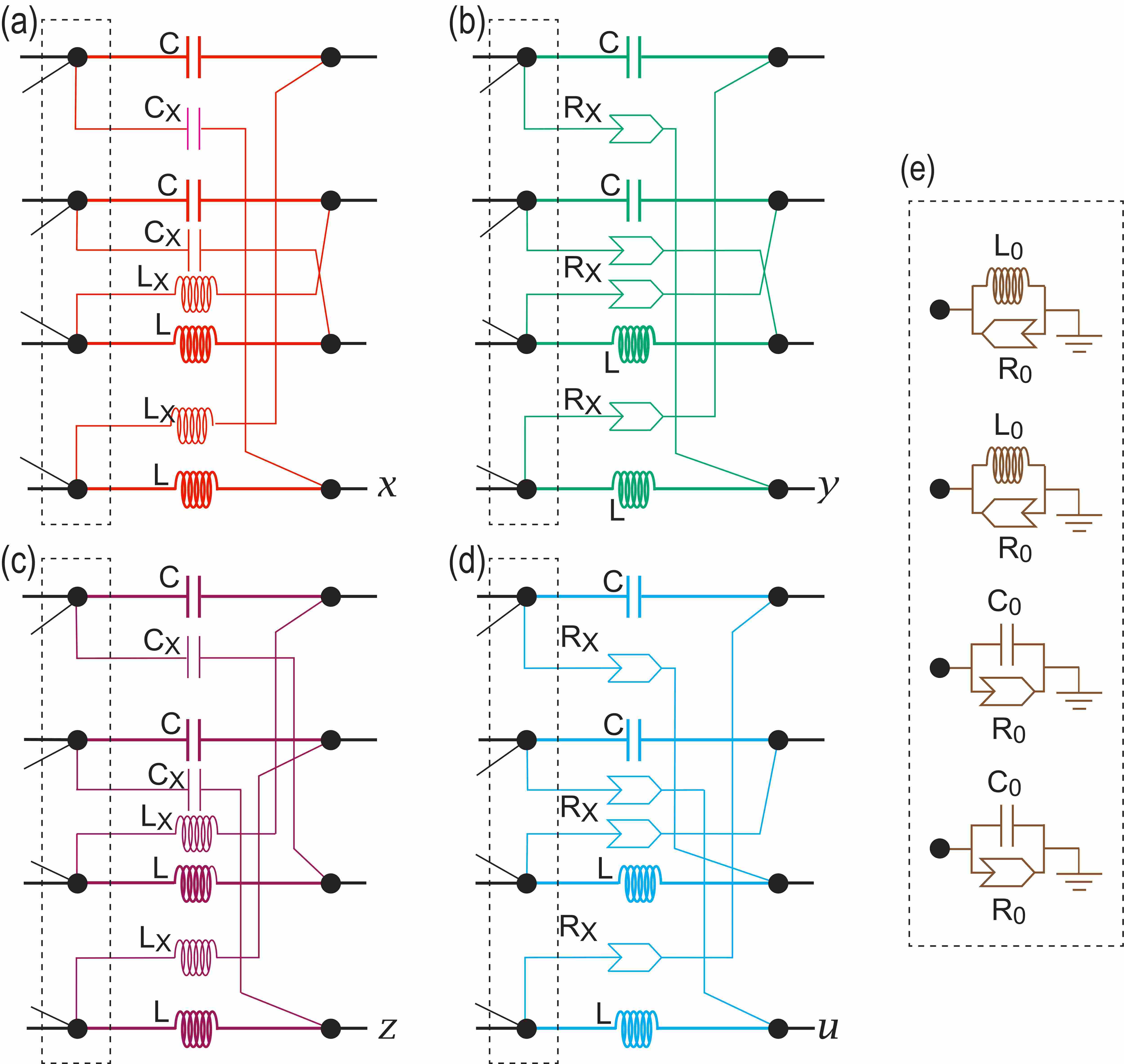}}
\caption{Basic elements for electric circuit realization of the second-Chern
insulator on the 4D hypercubic lattice. The dotted rectangul represents
the unit cell. Colored circuits in (a) red, (b) green, (c) purple and (d)
cyan represent links parallel to four independent axes $x$, $y$, $z$ and $u$, 
respectively. (e) Each node in the unit cell is grounded either by a
capacitance or an inductor with an operational amplifier. See also Fig.\protect\ref{FigProject} how to
construct a whole circuit using the unit cell and links. }
\label{FigCircuit}
\end{figure}

Second, we associate one unit cell with each site, and one link circuit with
each link: There are four types of link circuits in 4D lattice as
illustrated in Fig.\ref{FigCircuit}. Actual forms of the unit cell and the
link circuits are constructed by analyzing the Kirchhoff current law.

The Kirchhoff current law of the circuit under the application of an AC
voltage $V\left( t\right) =V\left( 0\right) e^{i\omega t}$ is given by\cite%
{ComPhys,TECNature}, 
\begin{equation}
I_{a}\left( \omega \right) =\sum_{b}\mathcal{J}_{ab}\left( \omega \right)
V_{b}\left( \omega \right) ,  \label{CircuLap}
\end{equation}%
where the sum is taken over all adjacent nodes $b$, and $\mathcal{J}_{ab}\left( \omega \right) $ 
is called the circuit Laplacian. The
eigenvalues of the circuit Laplacian are called the admittance spectrum,
which provides us with an information of the bulk spectrum of the
corresponding Hamiltonian.

We explain the method to implement an electric circuit corresponding to the
Hamiltonian. The Hamiltonian is written in the form of the $2n\times 2n$
matrix. Each component has the form of $e^{\pm ik_{\mu }}$, $-e^{\pm ik_{\mu
}}$or $ie^{\pm ik_{\mu }}$. We express the term $e^{\pm ik_{\mu }}$ by
capacitors $i\omega Ce^{\pm ik_{\mu }}$, the term $-e^{\pm ik_{\mu }}$ by
inductors $-\frac{1}{i\omega L}e^{\pm ik_{\mu }}$, and the term $ie^{\pm
ik_{\mu }}$ by operational amplifiers $iR_{X}e^{\pm ik_{\mu }}$ as in the
case of the previous studies\cite%
{ComPhys,TECNature,Garcia,Hel,Lu,EzawaTEC,Hofmann,EzawaChern,EzawaMajo}. We
introduce operational amplifiers $R_{0}$ in order to make the Dirac mass
complex. Then we connect each node by the capacitor $C_{0}$ and the inductor 
$L_{0}$ to the ground as in Fig.\ref{FigCircuit}(e).

Let us explicitly show the implementation in the case of 4Ds. In order to
make a circuit simple, we take a representation $\Gamma ^{0}=\sigma
_{z}\otimes I$, $\Gamma ^{x}=\sigma _{x}\otimes \sigma _{x}$, $\Gamma
^{y}=\sigma _{x}\otimes \sigma _{y}$, $\Gamma ^{z}=\sigma _{x}\otimes \sigma
_{z}$ and $\Gamma ^{u}=\sigma _{y}\otimes I$ for the second-Chern insulator.
We show the element of the circuit structure in Fig.\ref{FigCircuit}. We use
these structures colored in red, green, purple and cyan as $x$, $y$, $z$ and 
$u$ links in Fig.\ref{FigProject}, while the unit cell is denoted by black
dotted rectangles.

The circuit Laplacian is a sum of five terms in 4Ds, $\mathcal{J}=\sum_{\mu
=0,x,y,z,u}\mathcal{J}_{\mu }$, with%
\begin{eqnarray}
\mathcal{J}_{0} &=&\left( 
\begin{array}{cccc}
m^{+} & 0 & 0 & 0 \\ 
0 & m^{+} & 0 & 0 \\ 
0 & 0 & m^{-} & 0 \\ 
0 & 0 & 0 & m^{-}%
\end{array}%
\right) ,\quad \\
\mathcal{J}_{x} &=&\left( 
\begin{array}{cccc}
f_{x}^{+} & 0 & 0 & g_{x}^{0} \\ 
0 & f_{x}^{+} & g_{x}^{0} & 0 \\ 
0 & g_{x}^{0} & f_{x}^{-} & 0 \\ 
g_{x}^{0} & 0 & 0 & f_{x}^{-}%
\end{array}%
\right) , \\
\mathcal{J}_{y} &=&\left( 
\begin{array}{cccc}
f_{y}^{+} & 0 & 0 & g_{y}^{-} \\ 
0 & f_{y}^{+} & g_{y}^{+} & 0 \\ 
0 & g_{y}^{-} & f_{y}^{-} & 0 \\ 
g_{y}^{+} & 0 & 0 & f_{y}^{-}%
\end{array}%
\right) , \\
\mathcal{J}_{z} &=&\left( 
\begin{array}{cccc}
f_{z}^{+} & 0 & g_{z}^{0} & 0 \\ 
0 & f_{z}^{+} & 0 & -g_{z}^{0} \\ 
g_{z}^{0} & 0 & f_{z}^{-} & 0 \\ 
0 & -g_{z}^{0} & 0 & f_{z}^{-}%
\end{array}%
\right) , \\
\mathcal{J}_{u} &=&\left( 
\begin{array}{cccc}
f_{u}^{+} & 0 & g_{u}^{-} & 0 \\ 
0 & f_{u}^{+} & 0 & g_{u}^{-} \\ 
g_{u}^{+} & 0 & f_{u}^{-} & 0 \\ 
0 & g_{u}^{+} & 0 & f_{u}^{-}%
\end{array}%
\right)
\end{eqnarray}%
with%
\begin{align}
m^{+}& =C_{0}-(i\omega R_{0})^{-1}, \\
m^{-}& =-(\omega ^{2}L_{0})^{-1}+(i\omega R_{0})^{-1}, \\
f_{j}^{+}& =2C\cos k_{j}-2C, \\
f_{j}^{-}& =-2(\omega ^{2}L)^{-1}\cos k_{j}+2(\omega ^{2}L)^{-1}, \\
g_{j}^{+}& =-(\omega ^{2}L_{X})^{-1}e^{ik_{j}}+C_{X}e^{-ik_{j}},
\label{ffgg} \\
g_{j}^{-}& =C_{X}e^{ik_{j}}-(\omega ^{2}L_{X})^{-1}e^{-ik_{j}}, \\
g_{j}^{0}& =(i\omega R_{X})^{-1}e^{ik_{j}}-(i\omega R_{X})^{-1}e^{-ik_{j}}.
\end{align}%
The system becomes an $LC$ resonator.

Finally, we fix parameters in electric circuit so that the circuit Laplacian 
$\mathcal{J}$ is identical to the Hamiltonian $H$. It follows that $\mathcal{J}=i\omega H$, 
provided the resonance frequency is given by 
\begin{equation}
\omega _{0}=1/\sqrt{LC}=1/\sqrt{L_{0}C_{0}},
\end{equation}%
and 
\begin{eqnarray}
t &=&C,\lambda =C_{X}=(\omega _{0}R_{X})^{-1}, \\
m &=&-4nC+C_{0}-(i\omega R_{0})^{-1}.
\end{eqnarray}%
An important observation is that we may change the mass $m$ by tuning $C_{0}$. 
Then, we may induce topological phase transitions by controlling the Chern
number based on the formula (\ref{ChernK}) together with (\ref{d0}).

The eigenvalues of the circuit Laplacian give an admittance spectrum\cite%
{ComPhys,TECNature,Garcia,Hel,Lu,EzawaTEC}. Then the DOS of the Laplacian is
proportional to the DOS of the Hamiltonian. Thus, the DOS of the Laplacian
is measured by the DOS of the admittance.

\section{Discussions}

We have shown that topological insulators in any dimensions are simulated by
electric circuits. In elementary particle physics, the spatial dimension is
believed to be higher than three according to string theory. On the other
hand, the classification table of the topological insulator dictates the
existence of the topological systems in higher dimensions\cite%
{Schny,Ryu,KawabataST}. However, these higher dimensional systems are
impossible to approach experimentally. Our results will open a rout to study
higher dimensional physics in laboratory.

The author is very much grateful to N. Nagaosa for helpful discussions on
the subject. This work is supported by the Grants-in-Aid for Scientific
Research from MEXT KAKENHI (Grants No. JP17K05490, No. JP15H05854 and No.
JP18H03676). This work is also supported by CREST, JST (JPMJCR16F1 and
JPMJCR1874).

\end{document}